\documentclass[aps,prl,showpacs,twocolumn,superscriptaddress]{revtex4}%
\usepackage{graphics}
\usepackage{amsmath}
\usepackage{graphicx}
\usepackage{amsfonts}
\usepackage{amssymb}%
\begin{document}
\title{Non-Markovian Reactivation of Quantum Relays}

\begin{abstract}
We consider a quantum relay which is used by two parties to perform several
continuous-variable protocols: Entanglement swapping, distillation, quantum
teleportation, and quantum key distribution. The theory of these protocols is
extended to a non-Markovian model of decoherence characterized by correlated
Gaussian noise. Even if bipartite entanglement is completely lost at the
relay, we show that the various protocols can progressively be reactivated by
the separable noise-correlations of the environment. In fact, above a critical
amount, these correlations are able to restore the distribution of
quadripartite entanglement, which can be localized into an exploitable
bipartite form by the action of the relay. Our findings are confirmed by a
proof-of-principle experiment and show the potential advantages of
non-Markovian effects in a quantum network architecture.

\end{abstract}

\author{Stefano Pirandola}
\email{stefano.pirandola@york.ac.uk}
\affiliation{Computer Science and York Centre for Quantum Technologies, University of York,
York YO10 5GH, United Kingdom}
\author{Carlo Ottaviani}
\affiliation{Computer Science and York Centre for Quantum Technologies, University of York,
York YO10 5GH, United Kingdom}
\author{Christian S. Jacobsen}
\affiliation{Department of Physics, Technical University of Denmark, Fysikvej, 2800 Kongens
Lyngby, Denmark}
\author{Gaetana Spedalieri}
\affiliation{Computer Science and York Centre for Quantum Technologies, University of York,
York YO10 5GH, United Kingdom}
\author{Samuel L. Braunstein}
\affiliation{Computer Science and York Centre for Quantum Technologies, University of York,
York YO10 5GH, United Kingdom}
\author{Tobias Gehring}
\affiliation{Department of Physics, Technical University of Denmark, Fysikvej, 2800 Kongens
Lyngby, Denmark}
\author{Ulrik L. Andersen}
\affiliation{Department of Physics, Technical University of Denmark, Fysikvej, 2800 Kongens
Lyngby, Denmark}
\maketitle

The concept of a relay is at the basis of network information
theory~\cite{CoverThomas}. Indeed the simplest network topology is composed by
three nodes: two end-users, Alice and Bob, plus a third party, the relay,
which assists their communication. This scenario is inherited by quantum
information theory~\cite{Mwilde,NielsenBook}, where the mediation of a quantum
relay can be found in a series of fundamental protocols. By sending quantum
systems to a middle relay, Alice and Bob may perform entanglement
swapping~\cite{Zukowski,EntSwap,EntSwap2,GaussSWAP}, entanglement
distillation~\cite{Briegel}, quantum teleportation~\cite{Tele,Tele2} and
quantum key distribution (QKD)~\cite{mdiQKD,Untrusted}.

Quantum relays are crucial elements for quantum network architectures at any
scale, from short-range implementations on quantum chips to long-distance
quantum communication. In all cases, their working mechanism has been studied
assuming Markovian decoherence models, where the errors are independent and
identically distributed (iid). Removing this iid approximation is one of the
goals of modern quantum information theory.

In a quantum chip (e.g., photonic~\cite{photonic1,photonic2} or
superconducting~\cite{chip3}), quantum relays can distribute entanglement
among registers and teleport quantum gates. Miniaturizing this architecture,
correlated errors may come from unwanted interactions between quantum systems.
A common bath may be introduced by a variety of imperfections, e.g., due to
diffraction, slow electronics etc. It is important to realize that
non-Markovian dynamics~\cite{Petruccione} will become increasingly important
as the size of quantum chips further shrinks.

At long distances (in free-space or fibre), quantum relays intervene to assist
quantum communication, entanglement and key distribution. Here,
noise-correlations and memory effects may naturally arise when optical modes
are employed in high-speed communications~\cite{UlrikCORR}, or propagate
through atmospheric turbulence~\cite{Tyler09,Semenov09,Boyd11} and
diffraction-limited linear systems~\cite{Lupo1,Lupo2}. Most importantly,
correlated errors must be considered in relay-based QKD, where an eavesdropper
(Eve) may jointly attack the two links with the relay (random permutations and
de Finetti arguments~\cite{Renner1,Renner2} cannot remove these residual
correlations). Eve can manipulate the relay itself as assumed in
measurement-device independent QKD~\cite{mdiQKD,Untrusted}. Furthermore,
Alice's and Bob's setups may also be subject to correlated side-channel attacks.

For all these reasons, we generalize the study of quantum relays to
non-Markovian conditions, developing the theory for continuous variable (CV)
systems~\cite{RMP} (qubits are discussed in the Supplemental Material). We
consider an environment whose Gaussian noise may be correlated between the two
links. In this scenario, while the relay always performs the same measurement,
the parties may implement different protocols (swapping, distillation,
teleportation, or QKD) all based, directly or indirectly, on the exploitation
of bipartite entanglement.

We find a surprising behavior in conditions of extreme decoherence. We
consider entanglement-breaking links~\cite{EBchannels,HolevoEB}, so that no
protocol can work under Markovian conditions. We then induce non-Markovian
effects by progressively increasing the noise correlations in the environment
while keeping their nature separable (so that there is no external reservoir
of entanglement). While these correlations are not able to re-establish
bipartite entanglement (or tripartite entanglement) we find that a critical
amount reactivates quadripartite entanglement, between the setups and the
modes transmitted. In other words, by increasing the separable correlations
above a `reactivation threshold' we can retrieve the otherwise lost
quadripartite entanglement (it is in this sense that we talk of `reactivated'
entanglement below). The measurement of the relay can then localize this
multipartite entanglement into a bipartite form, shared by the two remote
parties and exploitable for the various protocols.

As a matter of fact, we find that all the quantum protocols can be
reactivated. In particular, their reactivation occurs in a progressive
fashion, so that increasing the environmental correlations first reactivates
entanglement swapping and teleportation, then entanglement distillation and
finally QKD. Our theory is confirmed by a proof-of-principle experiment which
shows the reactivation of the most nested protocol, i.e., the QKD\ protocol.

\section{Results}

\begin{figure}[ptbh]
\vspace{-1.5cm}
\par
\begin{center}
\includegraphics[width=0.50\textwidth] {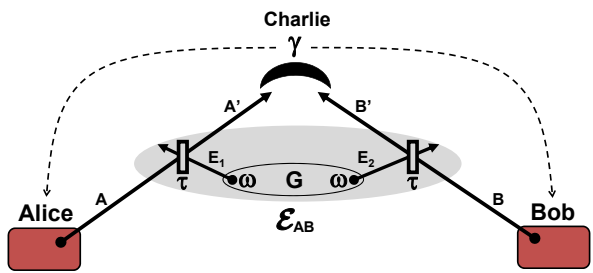}
\end{center}
\par
\vspace{-1.5cm}\caption{\textbf{Quantum relay.} Alice and Bob
connect their devices (red boxes) to a quantum relay, Charlie, for
implementing a CV protocol. On the received modes, Charlie always
performs a CV Bell detection whose outcome $\gamma$ is broadcast.
\textbf{Separable Gaussian environment.}
The travelling modes are subject to a joint Gaussian channel $\mathcal{E}%
_{AB}$. This is realized by two beam splitters with tranmissivity $\tau$ which
mix $A$ and $B$ with two ancillary modes, $E_{1}$ and $E_{2}$, respectively.
These ancillas inject thermal noise with variance $\omega$ and belong to a
correlated (but separable) Gaussian state $\rho_{E_{1}E_{2}}$.
\textbf{Entanglement breaking.} For $\omega>\omega_{\text{EB}}(\tau)$,
bipartite (and tripartite) entanglement cannot survive at the relay. In
particular, $A^{\prime}$ is disentangled from Alice's device, and $B^{\prime}$
is disentangled from Bob's, no matter if the environment is correlated or not.
\textbf{Non-Markovian reactivation.} Above a critical amount of separable
correlations, quadripartite entanglement is reactivated between Alice's and
Bob's devices and the transmitted modes, $A^{\prime}$ and $B^{\prime}$. Bell
detection can localize this multipartite resource into a bipartite form and
reactivate all the protocols.}%
\label{generalFIG}%
\end{figure}

\textit{General scenario}.--~As depicted in Fig.~\ref{generalFIG},
we consider two parties, Alice and Bob, whose devices are
connected to a quantum relay, Charlie, with the aim of
implementing a CV\ protocol (swapping, distillation,
teleportation, or QKD). The connection is established by sending
two modes, $A$ and $B$, through a joint quantum channel
$\mathcal{E}_{AB}$, whose outputs $A^{\prime}$ and $B^{\prime}$
are subject to a CV Bell detection~\cite{BellFORMULA}. This means
that modes $A^{\prime}$ and $B^{\prime}$ are mixed at a balanced
beam splitter and then homodyned, one in the position quadrature
$\hat{q}_{-}=(\hat{q}_{A^{\prime}}-\hat{q}_{B^{\prime
}})/\sqrt{2}$ and the other in the momentum quadrature
$\hat{p}_{+}=(\hat
{p}_{A^{\prime}}+\hat{p}_{B^{\prime}})/\sqrt{2}$. The classical
outcomes $q_{-}$ and $p_{+}$ can be combined into a complex
variable $\gamma :=q_{-}+ip_{+}$, which is broadcast to Alice and
Bob through a classical public channel. 

The joint quantum channel $\mathcal{E}_{AB}$ corresponds to an environment
with correlated Gaussian noise. This is modelled by two beam splitters (with
transmissivity $0<\tau<1$) mixing modes $A$ and $B$ with two ancillary modes,
$E_{1}$ and $E_{2}$, respectively (see Fig.~\ref{generalFIG}). These ancillas
are taken in a zero-mean Gaussian state~\cite{RMP} $\rho_{E_{1}E_{2}}$ with
covariance matrix (CM) in the symmetric normal form
\[
\mathbf{V}_{E_{1}E_{2}}(\omega,g,g^{\prime})=\left(
\begin{array}
[c]{cc}%
\omega\mathbf{I} & \mathbf{G}\\
\mathbf{G} & \omega\mathbf{I}%
\end{array}
\right)  ,~%
\begin{array}
[c]{c}%
\mathbf{I}:=\mathrm{diag}(1,1),\\
\mathbf{G}:=\mathrm{diag}(g,g^{\prime}).
\end{array}
\]
Here $\omega\geq1$ is the variance of local thermal noise, while the block
$\mathbf{G}$ accounts for noise-correlations.

For $\mathbf{G}=\mathbf{0}$ we retrieve the standard Markovian case, based on
two independent lossy channels~\cite{EntSwap,EntSwap2,GaussSWAP}. For
$\mathbf{G}\neq\mathbf{0}$, the lossy channels become correlated, and the
local dynamics cannot reproduce the global non-Markovian evolution of the
system. Such a separation becomes more evident by increasing the correlation
parameters, $g$ and $g^{\prime}$, whose values are bounded by the bona-fide
conditions $|g|<\omega$, $|g^{\prime}|<\omega$, and $\omega\left\vert
g+g^{\prime}\right\vert \leq\omega^{2}+gg^{\prime}-1$%
~\cite{TwomodePRA,NJPpirs}. In particular, we consider the realistic case of
separable environments ($\rho_{E_{1}E_{2}}$ separable), identified by the
additional constraint $\omega\left\vert g-g^{\prime}\right\vert \leq\omega
^{2}-gg^{\prime}-1$~\cite{NJPpirs}. The amount of separable correlations can
be quantified by the quantum mutual information~\cite{Mwilde} $I(g,g^{\prime
})$, which is the sum of the quantum discord~\cite{RMPdiscord,OptimalDiscord}
and the classical correlations (see Supplemental Material).

\begin{figure}[ptbh] \vspace{-0.64cm}
\par
\begin{center}
\includegraphics[width=0.35\textwidth]{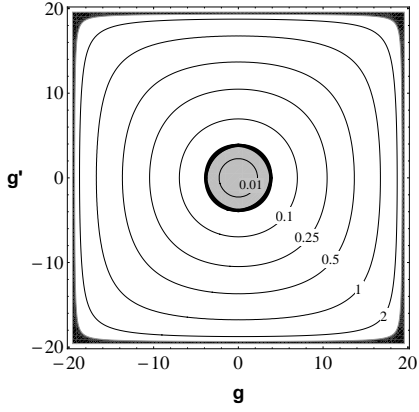}
\end{center}
\par
\vspace{-0.7cm}\caption{Non-Markovian reactivation of $1\times3$ quadripartite
entanglement. Assuming maximally-entangled states for the parties, and
entanglement-breaking conditions (here $\tau=0.9$ and $\omega=1.02\times
\omega_{\text{EB}}=19.38$), we show how quadripartite entanglement is
reactivated by increasing the separable correlations of the environment (bits
of quantum mutual information, which are constant over the concentric contour
lines). Inside the gray region there is no quadripartite entanglement with
respect to any $1\times3$ grouping of the four modes $abA^{\prime}B^{\prime}$.
Outside the gray region all the possible $1\times3$ groupings are entangled.
The external black region is excluded, as it corresponds to entangled or
unphysical environments.}%
\label{ENCcorr}%
\end{figure}

To analyse entanglement breaking, assume the asymptotic infinite-energy
scenario where Alice's (Bob's) device has a remote mode $a$ ($b$) which is
maximally entangled with $A$ ($B$). We then study the separability properties
of the global system composed by $a$, $b$, $A^{\prime}$ and $B^{\prime}$. In
the Markovian case ($\mathbf{G}=\mathbf{0}$), all forms of entanglement
(bipartite, tripartite~\cite{tripartite}, and
quadripartite~\cite{quadripartite}) are absent for $\omega>\omega_{\text{EB}%
}(\tau):=(1+\tau)/(1-\tau)$, so that no protocol can work. In the
non-Markovian case ($\mathbf{G}\neq\mathbf{0}$) the presence of
separable correlations does not restore bipartite or tripartite
entanglement when $\omega>\omega_{\text{EB}}(\tau)$. However, a
sufficient amount of these correlations is able to reactivate
$1\times3$ quadripartite entanglement~\cite{quadripartite}, in
particular, between mode $a$ and the set of modes
$bA^{\prime}B^{\prime}$. See Fig.~\ref{ENCcorr}.

Once quadripartite entanglement is available, the Bell detection
on modes $A^{\prime}$ and $B^{\prime}$ can localize it into a
bipartite form for modes $a$ and $b$. For this reason,
entanglement swapping and the other protocols can be reactivated
by sufficiently-strong separable correlations. In the following,
we discuss these results in detail for each specific protocol,
starting from the basic scheme of entanglement swapping. For each
protocol, we first generalize the theory to non-Markovian
decoherence, showing how the various performances are connected.
Then, we analyze the protocols under entanglement breaking
conditions.

\textit{Entanglement swapping.}--~The standard source of Gaussian entanglement
is the EPR state~\cite{RMP}. This is a two-mode Gaussian state with zero
mean-value and CM%
\[
\mathbf{V}(\mu)=\left(
\begin{array}
[c]{cc}%
\mu\mathbf{I} & \sqrt{\mu^{2}-1}\mathbf{Z}\\
\sqrt{\mu^{2}-1}\mathbf{Z} & \mu\mathbf{I}%
\end{array}
\right)  ,~\mathbf{Z}:=\mathrm{diag}(1,-1),
\]
where the variance $\mu\geq1$ quantifies its entanglement. Indeed the
log-negativity~\cite{logNEG,logNEG1,logNEG2} is strictly increasing in $\mu$:
It is zero for $\mu=1$ and tends to infinity for large $\mu$.

\begin{figure}[ptbh]
\vspace{-0.6cm}
\par
\begin{center}
\includegraphics[width=0.53\textwidth] {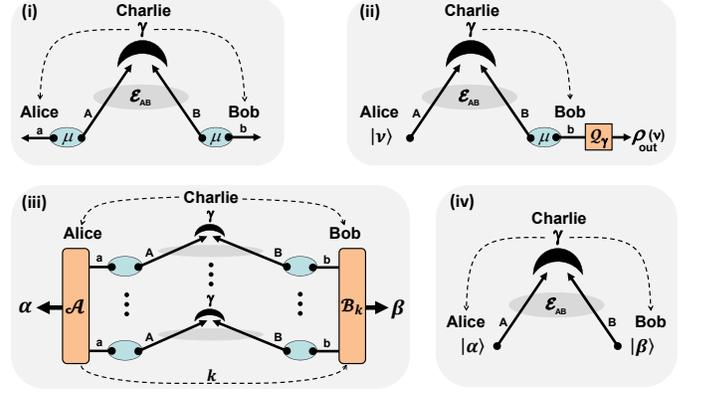}
\end{center}
\par
\vspace{-1.0cm}\caption{Relay-based quantum protocols in a correlated Gaussian
environment.\textbf{ (i)~Entanglement swapping.} Alice and Bob possess two EPR
states with variance $\mu$. Modes $A$ and $B$ are sent through the joint
channel $\mathcal{E}_{AB}$\ and received by Charlie. After the outcome
$\gamma$ is broadcast, the remote modes, $a$ and $b$, are projected into a
conditional state $\rho_{ab|\gamma}$. \textbf{(ii) Quantum teleportation}.
Alice's coherent state $\left\vert \nu\right\rangle $ is teleported into Bob's
state $\rho_{\text{out}}(\nu)$, after the communication of $\gamma$ and the
action of a conditional quantum operation $\mathcal{Q}_{\gamma}$.
\textbf{(iii) Entanglement/key distillation}. In the limit of many uses of the
relay, Alice performs a quantum instrument on her modes $a$, communicating a
classical variable $k$ to Bob, who performs a conditional quantum operation on
his modes $b$. This is a non-Gaussian quantum repeater where entanglement
swapping is followed by optimal one-way distillation.\textbf{ (iv) Practical
QKD}. Alice and Bob prepare Gaussian-modulated coherent states to be sent to
Charlie. The communication of the outcome $\gamma$ creates remote classical
correlations which are used to extract a secret key. Here the role of Charlie
could be played by Eve, so that the relay becomes an MDI-QKD\ node.}%
\label{swap}%
\end{figure}

Suppose that Alice and Bob have two identical EPR states, $\rho_{aA}(\mu)$
describing Alice's modes $a$ and $A$, and $\rho_{bB}(\mu)$ describing Bob's
modes $b$ and $B$, as in Fig.~\ref{swap}(i). They keep $a$ and $b$, while
sending $A$ and $B$ to Charlie through the joint channel $\mathcal{E}_{AB}$ of
the Gaussian environment. After the broadcast of the outcome $\gamma$, the
remote modes $a$ and $b$ are projected into a conditional Gaussian state
$\rho_{ab|\gamma}$, with mean-value $\mathbf{x}=\mathbf{x}(\gamma)$ and
conditional CM $\mathbf{V}_{ab|\gamma}$. In the Supplemental Material, we
compute
\begin{equation}
\mathbf{V}_{ab|\gamma}=\left(
\begin{array}
[c]{cc}%
\mathbf{A} & \mathbf{C}\\
\mathbf{C}^{T} & \mathbf{B}%
\end{array}
\right)  ,\label{CMmain}%
\end{equation}
where the $2\times2$ blocks are given by%
\begin{align}
\mathbf{A} &  =\mathbf{B}=\mathrm{diag}\left[  \mu-\frac{\mu^{2}-1}%
{2(\mu+\kappa)},\mu-\frac{\mu^{2}-1}{2(\mu+\kappa^{\prime})}\right]
,\label{blockBmain}\\
\mathbf{C} &  =\mathrm{diag}\left[  \frac{\mu^{2}-1}{2(\mu+\kappa)},-\frac
{\mu^{2}-1}{2(\mu+\kappa^{\prime})}\right]  ,\label{blockCmain}%
\end{align}
and the $\kappa$'s contain all the environmental parameters%
\begin{equation}
\kappa:=(\tau^{-1}-1)(\omega-g),~\kappa^{\prime}:=(\tau^{-1}-1)(\omega
+g^{\prime}).
\end{equation}

From $\mathbf{V}_{ab|\gamma}$ we compute the log-negativity $\mathcal{N}%
=\max\{0,-\log_{2}\varepsilon\}$ of the swapped state, in terms of the
smallest partially-transposed symplectic\ eigenvalue $\varepsilon$~\cite{RMP}.
In the Supplemental Material, we derive%
\begin{equation}
\varepsilon=\left[  \frac{(1+\mu\kappa)(1+\mu\kappa^{\prime})}{(\mu
+\kappa)(\mu+\kappa^{\prime})}\right]  ^{1/2}. \label{epsFINITI}%
\end{equation}
For any input entanglement ($\mu>1$), swapping is successful ($\varepsilon<1$)
whenever the environment has enough correlations to satisfy the condition
$\kappa\kappa^{\prime}<1$. The actual amount of swapped entanglement
$\mathcal{N}$ increases in $\mu$, reaching its asymptotic optimum for large
$\mu$, where
\[
\varepsilon\simeq\varepsilon_{\text{opt}}:=\sqrt{\kappa\kappa^{\prime}}.
\]

\textit{Quantum teleportation}.--~As depicted in Fig.~\ref{swap}(ii), we
consider Charlie acting as a teleporter of a coherent state $\left\vert
\nu\right\rangle $ from Alice to Bob. Alice's state and part of Bob's
EPR\ state are transmitted to Charlie through the joint channel $\mathcal{E}%
_{AB}$. After detection, the outcome $\gamma$ is communicated to Bob, who
performs a conditional quantum operation~\cite{NielsenBook} $\mathcal{Q}%
_{\gamma}$ on mode $b$ to retrieve the teleported state $\rho_{\text{out}}%
(\nu)\simeq\left\vert \nu\right\rangle \left\langle \nu\right\vert $. In the
Supplemental Material, we find a formula for the teleportation fidelity
$F=F(\mu,\kappa,\kappa^{\prime})$, which becomes asymptotically optimal for
large $\mu$, where%
\begin{equation}
F\simeq F_{\text{opt}}:=\left[  (1+\kappa)(1+\kappa^{\prime})\right]
^{-1/2}\leq(1+\varepsilon_{\text{opt}})^{-1}.\label{fidEPS}%
\end{equation}
Thus, there is a direct connection between the asymptotic protocols of
teleportation and swapping: If swapping fails ($\varepsilon_{\text{opt}}\geq
1$), teleportation is classical ($F_{\text{opt}}\leq1/2$~\cite{RMP}). We
retrieve the relation $F_{\text{opt}}=(1+\varepsilon_{\text{opt}})^{-1}$ in
environments with antisymmetric correlations $g+g^{\prime}=0$.

\textit{Entanglement distillation}.--~Entanglement distillation can be
operated on top of entanglement swapping as depicted in Fig.~\ref{swap}(iii).
After the parties have run the swapping protocol many times and stored their
remote modes in quantum memories, they can perform a one-way entanglement
distillation protocol on the whole set of swapped states $\rho_{ab|\gamma}$.
This consists of Alice locally applying an optimal quantum
instrument~\cite{Qinstrument} $\mathcal{A}$ on her modes $a$, whose quantum
outcome $\boldsymbol{\alpha}$\ is a distilled system while the classical
outcome $k$ is communicated. Upon receipt of $k$, Bob performs a conditional
quantum operation $\mathcal{B}_{k}$\ transforming his modes $b$ into a
distilled system $\boldsymbol{\beta}$.

The process can be designed in such a way that the distilled systems are
collapsed into entanglement bits (ebits), i.e., Bell state
pairs~\cite{NielsenBook}. The optimal distillation rate (ebits per relay use)
is lower-bounded~\cite{Qinstrument} by the coherent information
$I_{\mathcal{C}}$~\cite{CohINFO,CohINFO2} computed on the single copy state
$\rho_{ab|\gamma}$. In the Supplemental Material, we find a closed expression
$I_{\mathcal{C}}=I_{\mathcal{C}}(\mu,\kappa,\kappa^{\prime})$ which is
maximized for large $\mu$, where $I_{\mathcal{C}}\simeq-\log_{2}%
(e\varepsilon_{\text{opt}})$. Asymptotically, entanglement can be distilled
for $\varepsilon_{\text{opt}}<e^{-1}\simeq0.367$.

\textit{Secret key distillation}.--~The scheme of Fig.~\ref{swap}(iii) can be
modified into a key distillation protocol, where Charlie (or Eve~\cite{mdiQKD}%
) distributes secret correlations to Alice and Bob, while the environment is
the effect of a Gaussian attack. Alice's quantum instrument is here a
measurement with classical outputs $\boldsymbol{\alpha}$ (the secret key) and
$k$ (data for Bob). Bob's operation is a measurement conditioned on $k$, which
provides the classical output $\boldsymbol{\beta}$ (key estimate). This is an
ideal key-distribution protocol~\cite{KeyCAP} whose rate is lower-bounded by
the coherent information, i.e., $K\geq I_{\mathcal{C}}$~(see Supplemental Material).

\textit{Practical QKD}.--~The previous key-distribution\ protocol can be
simplified by removing quantum memories and using single-mode measurements, in
particular, heterodyne detections. This is equivalent to a run-by-run
preparation of coherent states, $\left\vert \alpha\right\rangle $ on Alice's
mode $A$, and $\left\vert \beta\right\rangle $ on Bob's mode $B$, whose
amplitudes are Gaussianly modulated with variance $\mu-1$. As shown in
Fig.~\ref{swap}(iv), these states are transmitted to Charlie (or
Eve~\cite{mdiQKD}) who measures and broadcasts $\gamma\simeq\alpha-\beta
^{\ast}$.

Assuming ideal reconciliation~\cite{RMP}, the secret key rate $R=R(\mu
,\kappa,\kappa^{\prime})$ increases in $\mu$. Modulation variances $\mu
\gtrsim50$ are experimentally achievable and well approximate the asymptotic
limit for $\mu\gg1$, where the key rate is optimal and satisfies~(see
Supplemental Material)%
\begin{equation}
R_{\text{opt}}\gtrsim\log_{2}\left(  \frac{F_{\text{opt}}}{e^{2}%
\varepsilon_{\text{opt}}}\right)  +h(1+2\varepsilon_{\text{opt}}),
\label{rateTEXT}%
\end{equation}
with $h(x):=\frac{x+1}{2}\log_{2}\frac{x+1}{2}-\frac{x-1}{2}\log_{2}\frac
{x-1}{2}$. Using Eq.~(\ref{fidEPS}), we see that the right hand side of
Eq.~(\ref{rateTEXT}) can be positive only for $\varepsilon_{\text{opt}%
}\lesssim0.192$. Thus the practical QKD protocol is the most difficult to
reactivate: Its reactivation implies that of entanglement/key distillation and
that of entanglement swapping. This is true not only asymptotically but also
at finite $\mu$ as we show below.

\begin{figure}[ptbh]
\vspace{-0.15cm}
\par
\begin{center}
\includegraphics[width=0.48\textwidth]{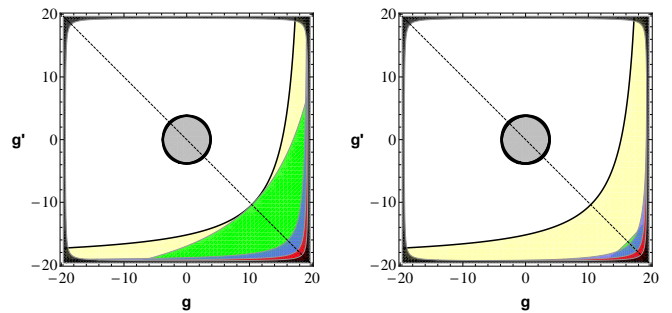}
\end{center}
\par
\vspace{-0.6cm}\caption{Non-Markovian reactivation of quantum protocols from
entanglement-breaking (here $\tau=0.9$ and $\omega=1.02\times\omega
_{\text{EB}}=19.38$). Each point of the correlation plane corresponds to a
Gaussian environment with separable correlations. In panel \textbf{a} we
consider the optimal scenario of large $\mu$ (asymptotic protocols). Once
quadripartite $1\times3$ entanglement has been reactivated (outside the gray
ring), we have the progressive reactivation of entanglement swapping
($\mathcal{N}>0$, yellow region), quantum teleportation of coherent states
($F>1/2$, green region), entanglement/key distillation ($I_{\mathcal{C}}$,
$K>0$, blue region) and practical QKD ($R>0$, red region). Panel \textbf{b} as
in \textbf{a} but refers to a realistic scenario with experimentally
achievable values of $\mu$. We consider $\mu\simeq6.5$%
~\cite{TopENT,TobiasSqueezing} as input entanglement for the
entanglement-based protocols, and $\mu\simeq50$ as modulation for the
practical QKD\ protocol. The reactivation phenomenon persists and can be
explored with current technology. Apart from teleportation, the other
thresholds undergo small modifications.}%
\label{total}%
\end{figure}

\textit{Reactivation from entanglement breaking}.--~Once the theory of the
previous protocols has been extended to non-Markovian decoherence, we can
study their reactivation from entanglement breaking conditions. Consider an
environment with transmissivity $\tau$ and entanglement-breaking thermal noise
$\omega>\omega_{\text{EB}}(\tau)$, so that no protocol can work for
$\mathbf{G}=\mathbf{0}$. By increasing the separable correlations in the
environment, not only can quadripartite entanglement be reactivated but, above
a certain threshold, it can also be localized into a bipartite form by the
relay's Bell detection. Once entanglement swapping is reactivated, all other
protocols can progressively be reactivated. As shown in Fig.~\ref{total},
there are regions of the correlation plane where entanglement can be swapped
($\mathcal{N}>0$), teleportation is quantum ($F>1/2$), entanglement and keys
can be distilled ($I_{\mathcal{C}}$, $K>0$), and practical QKD can be
performed ($R>0$). This occurs both for large and experimentally-achievable
values of $\mu$. 

Note that the reactivation is asymmetric in the plane only because of the
specific Bell detection adopted, which generates correlations of the type
$g>0$ and $g^{\prime}<0$. Using another Bell detection (projecting onto
$\hat{q}_{+}$ and $\hat{p}_{-}$), the performances would be inverted with
respect to the origin of the plane. Furthermore, the entanglement localization
(i.e., the reactivation of entanglement swapping) is triggered for
correlations higher than those required for restoring quadripartite
entanglement, suggesting that there might exist a better quantum measurement
for this task. The performances of the various protocols improve by increasing
the separable correlations of the environment, with the fastest reactivation
being achieved along the diagonal $g+g^{\prime}=0$, where swapping and
teleportation are first recovered, then entanglement/key distillation and
practical QKD, which is the most nested region.

\textit{Correlated additive noise}.--~The phenomenon can also be found in
other types of non-Markovian Gaussian environments. Consider the limit for
$\tau\rightarrow1$ and $\omega\rightarrow+\infty$, while keeping constant
$n:=(1-\tau)\omega$, $c:=g(\omega-1)^{-1}$ and $c^{\prime}:=g^{\prime}%
(\omega-1)^{-1}$. This is an asymptotic environment which adds correlated
classical noise to modes $A$ and $B$, so that their quadratures undergo the
transformations%
\[
\left(  \hat{q}_{A},\hat{p}_{A},\hat{q}_{B},\hat{p}_{B}\right)  \rightarrow
\left(  \hat{q}_{A},\hat{p}_{A},\hat{q}_{B},\hat{p}_{B}\right)  +(\xi_{1}%
,\xi_{2},\xi_{3},\xi_{4}).
\]
Here the $\xi_{i}$'s are zero-mean Gaussian variables\ whose covariances
$\left\langle \xi_{i}\xi_{j}\right\rangle $ are specified by the classical CM%
\begin{equation}
\mathbf{V}\left(  n,c,c^{\prime}\right)  =n\left(
\begin{array}
[c]{cc}%
\mathbf{I} & \mathrm{diag}(c,c^{\prime})\\
\mathrm{diag}(c,c^{\prime}) & \mathbf{I}%
\end{array}
\right)  , \label{CMadditive}%
\end{equation}
where $n\geq0$ is the variance of the additive noise, and $-1\leq c,c^{\prime
}\leq1$ quantify the classical correlations. The entanglement-breaking
condition becomes $n>2$.

\begin{figure}[ptbh]
\vspace{+0.13cm}
\par
\begin{center}
\includegraphics[width=0.46\textwidth]{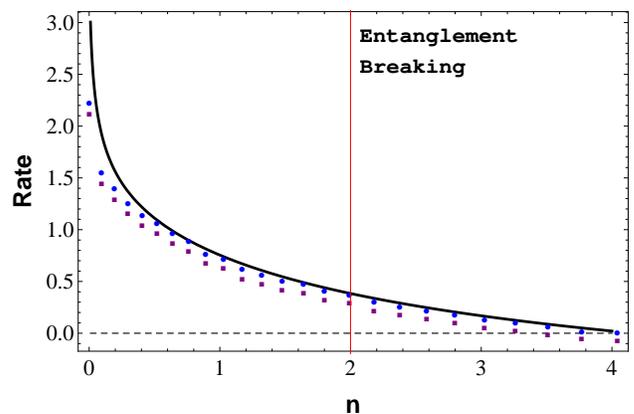}
\end{center}
\par
\vspace{-0.4cm}\caption{Plot the secret-key rate $R$ (bits per
relay use) as a function of the additive noise $n$. The solid
curve is the theoretical rate computed for a correlated-additive
environment ($c=c^{\prime}=1$) and\ realistic signal modulation
($\mu\simeq52$). This rate is shown to be positive after
entanglement breaking ($n>2$).\textbf{ }Points are experimental
data: Blue circles refer to ideal reconciliation, and purple
squares to achievable reconcilation efficiency ($\simeq0.97$). Due
to loss at the untrusted relay, the experimental key rate is
slightly below the theoretical curve (associated with the
correlated side-channel attack). The reactivation
of QKD is confirmed experimentally.}%
\label{ExpOUT}%
\end{figure}

To show non-Markovian effects, we consider the protocol which is the most
difficult to reactivate, the practical QKD protocol. We can specify its key
rate $R(\mu,n,c,c^{\prime})$ for $c=c^{\prime}=1$ and assume a realistic
modulation $\mu\simeq52$. We then plot $R$ as a function of the additive noise
$n$ in Fig.~\ref{ExpOUT}. As we can see, the rate decreases in $n$ but remains
positive in the region $2<n\leq4$ where the links with the relay become
entanglement-breaking. As we show below, this behaviour persists in the
presence of loss, as typically introduced by experimental imperfections.

\begin{figure}[ptbh]
\vspace{+0.0cm}
\par
\begin{center}
\includegraphics[width=0.45\textwidth]{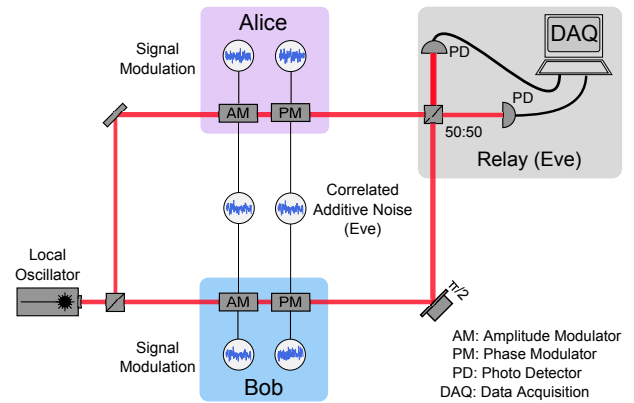}
\end{center}
\par
\vspace{-0.1cm}\caption{\textbf{Experimental setup}. Alice and Bob
receive 1064 nm light from the same laser source (local
oscillator). At both stations, the incoming beams are Gaussianly
modulated in phase and amplitude using electro-optical modulators
driven by uncorrelated signal generators. In addition, the phase
and amplitude modulators for Alice and Bob have correlated inputs
respectively, such that a noisy modulation identical for both
Alice and Bob is added to the phase and amplitude signals
(side-channel attack). The magnitudes of the correlated noise
modulations are progressively increased (from $n=0$ to $4$), and
kept symmetrical between the quadratures, while the signal
modulations are kept constant at the same level in both
quadratures for Alice and Bob ($\mu\simeq52$). At the untrusted
relay, the modes are mixed at a balanced beam splitter and the
output ports photo-detected, with an overall efficiency of
$\simeq98\%$. Photocurrents are then processed to realize a CV
Bell measurement. See Supplemental Material for details.}%
\label{SetupFIG}%
\end{figure}

\textit{Experimental results}.--~Our theoretical results are confirmed by a
proof-of-principle experiment, whose setup is schematically depicted in
Fig.~\ref{SetupFIG}. We consider Alice and Bob generating Gaussianly modulated
coherent states by means of independent electro-optical modulators, applied to
a common local oscillator. Simultaneously, the modulators are subject to a
side-channel attack: Additional electrical inputs are introduced by Eve, whose
effect is to generate additional and unknown phase-space displacements. In
particular, Eve's electrical inputs are correlated so that the resulting
optical displacements introduce a correlated-additive Gaussian environment
described by Eq.~(\ref{CMadditive}) with $c\simeq1$ and $c^{\prime}\simeq1$.
The optical modes then reach the midway relay, where they are mixed at a
balanced beam splitter and the output ports photo-detected. Although the
measurement is highly efficient, it introduces a small loss ($\simeq2\%$)
which is assumed to be exploited by Eve in the worst-case
scenario.

From the point of view of Alice and Bob, the side-channel attack and the
additional (small)\ loss at the relay are jointly perceived as a global
coherent Gaussian attack of the optical modes. Analysing the statistics of the
shared classical data and assuming that Eve controls the entire environmental
purification compatible with this data, the two parties may compute the
experimental secret-key rate (see details in the Supplemental Material). As we
can see from Fig.~\ref{ExpOUT}, the experimental points are slightly below the
theoretical curve associated with the correlated-additive environment,
reflecting the fact that the additional loss at the relay tends to degrade the
performance of the protocol. Remarkably, the experimental rate remains
positive after the entanglement-breaking threshold, so that\ the non-Markovian
reactivation of QKD is experimentally confirmed.

\section{Discussion}

We have theoretically and experimentally demonstrated that the most important
protocols operated by quantum relays can work in conditions of extreme
decoherence thanks to non-Markovian effects. Assuming high Gaussian noise in
the links, we have considered a regime where any form of entanglement
(bipartite, tripartite or quadripartite) is broken under Markovian conditions.
In this regime, we have proven that a suitable amount of separable
correlations can reactivate the distribution of $1\times3$ quadripartite
entanglement, and this resource can be successfully localised into a bipartite
form exploitable by Alice and Bob. As a result, all the basic protocols can be
progressively reactivated by the action of the relay.

The non-Markovian reactivation of a quantum relay is a new physical phenomenon
which points out several interesting facts. It shows that, in the absence of
bipartite entanglement, we can still rely on a multipartite form of this
resource, which can be manipulated, converted and exploited for quantum
information tasks. Then, it also shows how separable correlations can play a
role in the distribution of this multipartite resource within a quantum
network topology (see Supplemental Material for more details and relations
with previous literature).

In conclusion, our results show new perspectives for all quantum systems where
correlated errors, memory effects, and non-Markovian dynamics are the most
important form of decoherence. This may involve both very short-distance
implementations, such as chip-based quantum computing, and long-distance
implementations, as is the case of diffraction-limited quantum communication
or relay-based QKD, where the most general eavesdropping strategies are based
on correlated attacks. Thanks to their potential benefits, non-Markovian
effects should be regarded as a physical resource to be exploited in quantum
network implementations.

\section{Methods}

Theoretical and experimental methods are given in the Supplemental Material.
Theoretical methods contain details about the following points: (i)~Study of
the Gaussian environment with correlated thermal noise, including a full
analysis of its classical and quantum correlations. (ii)~Study of the various
forms of entanglement available before the Bell detection of the relay.
(iii)~Study of the entanglement swapping protocol, i.e., the computation of
the CM $\mathbf{V}_{ab|\gamma}$ in Eq.~(\ref{CMmain}) and the derivation of
the eigenvalue $\varepsilon$ in Eq.~(\ref{epsFINITI}). (iv)~Generalization of
the teleportation protocol with details on Bob's quantum operation
$\mathcal{Q}_{\gamma}$ and the analytical formula for the fidelity
$F(\mu,\kappa,\kappa^{\prime})$. (v)~Details of the distillation protocol with
the analytical formula of $I_{\mathcal{C}}(\mu,\kappa,\kappa^{\prime})$.
(vi)~Details of the ideal key-distillation protocol, discussion on
MDI-security, and proof of the lower-bound $K\geq I_{\mathcal{C}}$.
(vii)~Derivation of the general secret-key rate $R(\xi,\mu,\kappa
,\kappa^{\prime})$ of the practical QKD protocol, assuming arbitrary
reconciliation efficiency $\xi$ and modulation variance $\mu$. (viii)~Explicit
derivation of the optimal rate $R_{\text{opt}}$ and the proof of the tight
lower bound in Eq.~(\ref{rateTEXT}). (ix)~Derivation of the
correlated-additive environment as a limit of the correlated-thermal one.
(x)~Study of entanglement swapping and practical QKD in the
correlated-additive environment, providing the formula of the secret-key rate
$R(\xi,\mu,n,c,c^{\prime})$.

\section*{Acknowledgements}

S. P. acknowledges support from the EPSRC via the `Quantum Communications HUB'
(EP/M013472/1) and the grant `qDATA' (EP/L011298/1). S. P. also thanks the
Leverhulme Trust (research fellowship `qBIO'). T. G. acknowledges support from
the H. C. {\O }rsted postdoc programme. U. L. A. thanks the Danish Agency for
Science, Technology and Innovation (Sapere Aude project).

\end{document}